\documentclass[aps,prb,twocolumn,superscriptaddress]{revtex4-1}
\usepackage{amsmath,amssymb,color}

\usepackage{graphicx,xspace,textcomp,dcolumn}
\newcommand{\tc}{$T_c$\xspace}
\newcommand{\tn}{$T_N$\xspace}

\newcommand{\slrrt}     {$(T_1T)^{-1}$\xspace}

\newcommand{\laf}{LaFeAsO${_{1-x}}$F${_{x}}$\xspace}
\newcommand{\lao}{LaFeAsO\xspace}

\newcommand{\Rao}{RFeAsO\xspace}
\newcommand{\bkfa}{Ba$\mathrm{_{1-x}}$K$\mathrm{_x}$Fe$\mathrm{_2}$As$\mathrm{_2}$\xspace}
\newcommand{\bfa}{BaFe$\mathrm{_2}$As$\mathrm{_2}$\xspace}
\newcommand{\feas}{Fe$\mathrm{_2}$As$\mathrm{_2}$\xspace}
\newcommand{\bfca}{BaFe$\mathrm{_{2-x}}$Co$\mathrm{_x}$As$\mathrm{_2}$\xspace}

\newcommand{\Afa}{AFe$\mathrm{_2}$As$\mathrm{_2}$\xspace}

\newcommand{\lifeas}{{LiFeAs}\xspace}

\begin{document}

\title{Nematicity in LaFeAsO$_\mathrm{1-x}$F$_\mathrm{{x}}$}


\author{C. Hess}
\email{c.hess@ifw-dresden.de}
\affiliation{Leibniz-Institute for Solid State and Materials Research, IFW-Dresden, 01069 Dresden, Germany}
\affiliation{Center for Transport and Devices, TU Dresden, 01069 Dresden, Germany}

\author{H. Grafe}
\author{A. Kondrat}
\affiliation{Leibniz-Institute for Solid State and Materials Research, IFW-Dresden, 01069 Dresden, Germany}

\author{G. Lang}
\affiliation{Leibniz-Institute for Solid State and Materials Research, IFW-Dresden, 01069 Dresden, Germany}
\affiliation{ESPCI, PSL Research Univ.; CNRS; Sorbonne Univ., UPMC; LPEM, 10 rue Vauquelin, 75005 Paris, France}

\author{F. Hammerath}
\affiliation{Leibniz-Institute for Solid State and Materials Research, IFW-Dresden, 01069 Dresden, Germany}
\affiliation{Institute for Solid State Physics, TU Dresden, 01069 Dresden, Germany}

\author{L. Wang}
\altaffiliation{Institute for Solid-State Physics, Karlsruhe Institute of Technology, 76021 Karlsruhe, Germany}
\affiliation{Leibniz-Institute for Solid State and Materials Research, IFW-Dresden, 01069 Dresden, Germany}

\author{R. Klingeler}
\affiliation{Kirchhoff Institute of Physics, Heidelberg University, 69120 Heidelberg, Germany}

\author{G. Behr}
\thanks{Deceased}
\affiliation{Leibniz-Institute for Solid State and Materials Research, IFW-Dresden, 01069 Dresden, Germany}

\author{B. B\"uchner}
\affiliation{Leibniz-Institute for Solid State and Materials Research, IFW-Dresden, 01069 Dresden, Germany}
\affiliation{Center for Transport and Devices, TU Dresden, 01069 Dresden, Germany}
\affiliation{Institute for Solid State Physics, TU Dresden, 01069 Dresden, Germany}

\date{\today}

\begin{abstract}
Orbital ordering has recently  emerged as another important state in iron based superconductors, and its role for superconductivity as well as its connection to magnetic order and orthorhombic lattice distortion are heavily debated. In order to search for signatures of this so-called nematic phase in oxypnictides, we revisit the normal state properties of the pnictide superconductor \laf with a focus on resistivity, Nernst effect, thermal expansion, and $^{75}$As NMR data. The transport properties at the underdoped level $x=0.05$ exhibit pronounced anomalies at about the same temperature where undoped \lao develops long-range nematic ordering, i.e. at about 160~K. Furthermore, the $^{75}$As-NMR spin-lattice relaxation rate \slrrt reveals a progressive slowing down of spin fluctuations. Yet,  long-range magnetic order and also a detectable orthorhombic lattice distortion are absent. Thus, we conclude from the data that short-range orbital-nematic ordering or a slowly fluctuating form of it sets in near 160~K. Remarkably, all anomalies in the transport and also the indications of slow spin fluctuations disappear close to optimal doping $x=0.1$ which suggests that in \laf the nematic phase actually competes with superconductivity.
\end{abstract}

\maketitle

\section{Introduction}

The discovery of superconductivity in \laf \cite{Kamihara2008} initiated a tremendous research effort which yielded soon after a large variety of new superconducting iron pnictide compounds with \tc up to 55~K \cite{Johnston2010}. All compounds feature \feas-layers as the common structural unit, with typical examples being \Rao (R=La or Rare Earth), \Afa (A=alkaline earth or Eu), \lifeas and FeSe, which commonly are referred to as '1111', '122', '111' and '11' compounds, respectively (note that Se replaces As in the last example). While the latter two compounds exhibit superconductivity already in their stoichiometric form, the parent materials \Rao and \Afa are poor metals which exhibit an antiferromagnetic spin density wave (SDW) ground state. The temperature dependent transition to this state is always accompanied by a tetragonal-to-orthorhombic transition either at the same or at somewhat higher temperature. Chemical doping destabilizes this state in favor of superconductivity. The obvious proximity of superconductivity and antiferromagnetism very early has lead to the conjecture that spin fluctuations are the driving mechanism of superconductivity \cite{Mazin2008}. It has been noticed \cite{Xu2008,Fang2008,Fernandes2014} that the structural transition in the parent compounds is unlikely due to steric effects, i.e. phonon-driven, but rather electronically driven. In fact, evidence for additional orbital ordering occurring concomitantly with the structural distortion has recently inferred e.g. from resistivity anisotropy \cite{Chu2010,Chu2012} and angular resolved photoemission spectroscopy (ARPES) \cite{Yi2011}.
Thus, there are three types of long-range order which occur upon entering the low-temperature orthorhombic SDW state, viz. orbital ordering, the orthorhombic distortion, and the antiferromagnetic SDW order. All these types of ordering are coupled with each other and break $C_4$-symmetry and it is an ongoing question which of the three order parameters is driving the transition to the low-temperature phase that has been dubbed ``nematic'' phase, and to which extent it supports or competes with superconductivity \cite{Fernandes2014}.

The most comprehensive investigations of nematic order and fluctuations of it concern 122 and 11 phases \cite{Chu2010,Chu2012,Yi2011,Fernandes2013,Boehmer2014a,Baek2014} whereas relatively little is known for the 1111 phases. Viable routes to reach superconductivity in \Rao through chemical doping are the substitution of interlayer O by F \cite{Kamihara2008} and the substitution of Co for Fe within the \feas layers in both 1111 and 122 compounds \cite{Sefat2008b}. A generic observation upon doping is that the SDW phase is destabilized, i.e., the $T_s$ and $T_N$ gradually decrease and at some finite doping level superconductivity emerges.
The actual nature of the transition from SDW to superconductivity is much under debate. There is evidence that in \laf, on which we focus in this paper, the transition is abrupt and first order-like towards a homogeneous superconducting phase \cite{Luetkens2009,Hess2009} while in all other systems experiments suggest a finite doping interval where superconductivity and static magnetism coexist \cite{Drew2009}.

In the parent compound \lao the structural tetragonal-to-orthorhombic transition takes place at $T_s\approx160$~K \cite{Luetkens2009,Kondrat2009} and the long range antiferromagnetic SDW order sets in at $T_N\approx137$~K \cite{Luetkens2009,Cruz2008,Klauss2008,Klingeler2010}.
These successive phase transitions can be well detected due to pronounced signatures in transport, thermodynamic, magnetic, and structural quantities.
In this paper we compare transport data of \laf from resistivity $\rho$ \cite{Hess2009} and the Nernst coefficient $\nu$ \cite{Kondrat2011} with thermal expansion \cite{Wang2009a} and NMR \cite{Hammerath2013} data obtained on the same samples at selected doping levels $x=0$, 0.05, and 0.1.
We observe that distinct anomalies, present in both resistivity and the Nernst coefficient at the underdoped level $x=0.05$, occur at roughly the same temperature ($\sim$160~K) as the nematic ordering temperature in the parent compound. The $^{75}$As spin-lattice relaxation rate \slrrt shows a strong increase upon cooling, signaling the slowing-down of spin fluctuations \cite{Hammerath2013}, but long-range magnetic order as well as a 
global
orthorhombic distortion are never established \cite{Luetkens2009,Kondrat2009,Klingeler2010,Qureshi2010}.

These qualitative features disappear close to optimal doping $x=0.1$. We conclude from these findings that  short-range or fluctuating orbital ordering is still present in underdoped superconducting \laf but disappears when optimal superconductivity is reached. The results suggest further that in \laf the nematic phase competes with superconductivity.

\section{Experimental}

Sample preparation and all experimental work has been reported separately in earlier works \cite{Kondrat2009,Hess2009,Wang2009a,Hammerath2013}, and experimental details can be inferred from the respective publications.

\section{Results}

\subsection{The parent compound LaOFeAs}

\begin{figure}[t]

\centering

\includegraphics[width=0.85\columnwidth]{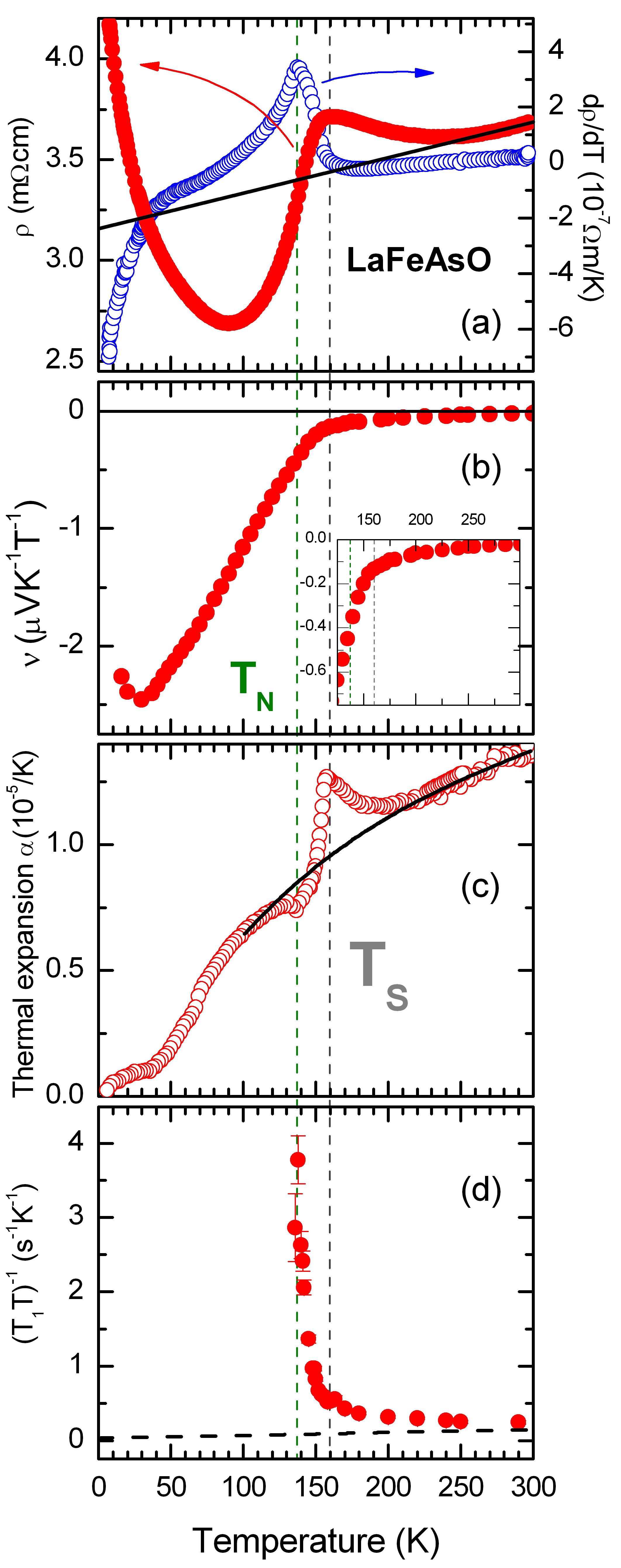}

\caption{Normalized resistivity $\rho(T)$ and the derivative $d\rho/dT$ \cite{Hess2009} (a), the Nernst coefficient $\nu(T)$ \cite{Kondrat2011} (b) thermal expansion \cite{Wang2009a} (c) and \slrrt \cite{Hammerath2013} (d) of \lao. Solid and dashed lines are guides to the eye.} \label{rho_undoped}

\end{figure}

The electrical resistivity, $\rho(T)$, of \lao (see figure~\ref{rho_undoped}a) develops a deviation from a standard metallic linear $T$-de\-pen\-dence near 300~K upon cooling which leads to a maximum at $T_s$ and a subsequent sharp drop with an inflection point at $T_N$ \cite{Kamihara2008,Hess2009,Klauss2008,Mcguire2008}.
A further decrease of temperature $T$ leads to a minimum of $\rho(T)$ at $\sim90$~K followed by a strong low-$T$ upturn.
The observed anomalies in the resistivity upon passing through $T_s$ and $T_N$ coarsely represent a canonical behavior of the resistivity upon entering the SDW ground state not only of many other iron pnictides  such as \bfa but also other very different SDW compounds like as Mn$_3$Si \cite{Steckel2014} or Yb$_2$Pt$_2$Pb \cite{Kim2008b}.
In \laf, due to the complex interplay of orbital, magnetic and structural degrees of freedom, the situation is, however, a bit more involved as compared to the transition to a presumably pure SDW state in the latter two compounds.
The detailed origin of such anomalous transport properties is currently not clear. Qualitatively, it seems straightforward, however, to rationalize the observed anomalies in terms of enhanced scattering at $T>T_s$,  presumably arising from some form of nematic fluctuations, and a reduced carrier density together with a dramatically reduced carrier scattering rate in the long-range orbital ordered orthorhombic state. In particular, the dramatic drop of $\rho(T)$ below $T_s$ implies a strong enhancement of the carrier relaxation time. The inflection point upon cooling through $T_N$ signals that the decrease of the resistivity becomes weaker in the magnetically ordered state, presumably due to the additional opening of an SDW gap which further reduces the carrier density.

The actual nature of the nematic fluctuations which give rise to the enhanced $\rho(T)$ at $T>T_s$ is uncertain. However, there is strong evidence that  fluctuations of orbital, lattice and spin degrees of freedom are ubiquitous which all potentially couple to the resistivity: On the one hand, the Nernst coefficient $\nu(T)$ \cite{Kondrat2011} as shown in Fig.~\ref{rho_undoped}b) not only exhibits a dramatic enhancement upon cooling through $T_s$, there is furthermore a clear fluctuation regime of an already enhanced $\nu(T)$ that extends almost up to room temperature, see inset of Fig.\ref{rho_undoped}b).
This quantity is known as a sensitive probe for reconstructions of the Fermi surface due to electronic-nematic or SDW order \cite{Steckel2014,Hess2010,Bel2003,Cyr-Choiniere2009,Daou2010,Hackl2009,Hackl2010}. Pure lattice effects typically are  of secondary importance.
Previously, the ``giant'' enhancement of the Nernst coefficient in \lao below $T_s$ has quite generally been attributed to the Fermi surface reconstructions due to the SDW state \cite{Kondrat2011}. However, the fact that the enhancement clearly sets in close to $T_s$, i.e., above the magnetic ordering temperature $T_s$, with practically no specific anomaly at $T_N$, implies the Nernst effect to respond primarily to orbital ordering.
This suggests the fluctuation regime of $\nu(T)$ at $T>T_s$, to be rather specifically connected to orbital fluctuations.

On the other hand, an extended \textit{lattice} fluctuation regime is observed in thermal expansion data at $T>T_s$ \cite{Wang2009a}, see Fig.~\ref{rho_undoped}c). 
These anomalous contributions to $\alpha(T)$ can be interpreted as a signature of elastic lattice softening. In the setup applied here where the samples experience a moderate compression along the direction of the detected length changes \cite{dila},
elastic softening is associated with shrinking of the sample showing up as a positive anomaly in $\alpha(T)$. The observed extended regime of anomalous length changes well above \tn\ hence probes the elastic soft mode which in case of an electronic origin of the structural transition can be associated with the nematic suceptibility~\cite{Chu2012,Fernandes2013}. Measurements of the elastic shear modulus performed in a capacitance dilatometer have indeed been used to obtain the nematic susceptibility of electron- and hole-doped \bfa\ and FeSe~\cite{Boehmer2014a,Boehmer2015}.

The presence of lattice fluctuations well above \tn\  is corroborated by the predominatly phononic heat conductivity $\kappa$ of \lao which exhibits a strong dip-like anomaly at $T\gtrsim T_s$ \cite{Kondrat2009,Mcguire2008} which also signals that structural fluctuations are relevant.
Finally, \textit{magnetic} fluctuations at least up to about 200~K can be clearly inferred from the $^{75}$As-NMR spin lattice relaxation rate, \slrrt, which diverges at $T_N$ \cite{Nakai2008} (see figure~\ref{rho_undoped}d).
Since the NMR relaxation rate is proportional \cite{Moriya1963} to the imaginary part of the dynamic spin susceptibility at the Lamor frequency, $(T_1T)^{-1}\sim \chi{''}({\bf q},\omega_0)$, (with the Larmor frequency $\omega_0$ and the wave vector $\bf q$), this increase shows the slowing down of spin fluctuations prior to the SDW ordering.

\subsection{Underdoped LaO$_{1-x}$F$_x$FeAs}

\begin{figure}[t]

\centering

\includegraphics[width=\columnwidth]{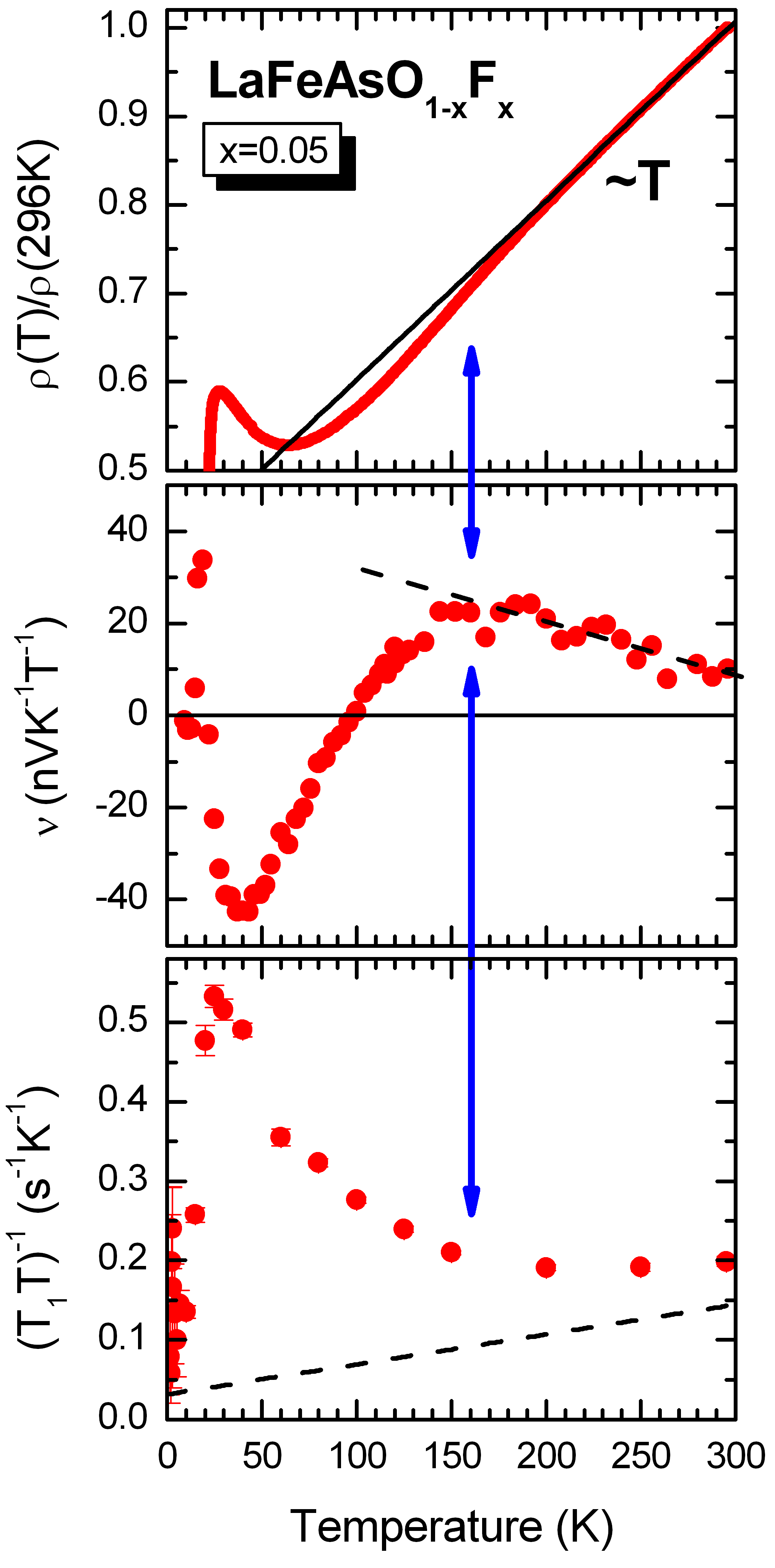}

\caption{Normalized resistivity $\rho(T)$ \cite{Hess2009} (a), Nernst coefficient $\nu(T)$ \cite{Kondrat2011} (b), and $^{75}$As-NMR spin-lattice relaxation rate \slrrt \cite{Hammerath2013} of \laf at $x=0.05$. Solid and dashed lines in (a) and (b) guides to the eye. Dashed line in (c): see text.} \label{all_five}

\end{figure}

It is enlightening to directly compare the above observations for the parent compound \lao with data at finite doping levels.
Figure~\ref{all_five} presents our data for $\rho(T)$, $\nu(T)$ and \slrrt\ of \laf at the underdoped level $x=0.05$, close to the ``border'' to the long-range ordered nematic phase.
As can be inferred very clearly from the data, the two transport quantities exhibit very clear anomalies at practically the same temperature ($T\sim160$~K) as in the parent compound: The dramatic drop of $\rho(T)$ at about $160$~K seen in the parent compound has been replaced by a much weaker but distinct anomaly (Fig.~\ref{all_five}a)): $\rho(T)$ drops below the low-$T$ extrapolation of its linear high temperature behavior at about 160~K, which suggests, as in the parent compound, a decrease of the transport relaxation rate \cite{Hess2009}.
Note, that prior to the onset of $T_c$ an upturn is present which indicates the localization of charge carriers.

The Nernst coefficient shown in Fig.~\ref{all_five}b) exhibits a very clear change of slope at $T\sim160$~K and a dramatic drop including a sign change upon cooling. In view of the above interpretation of the Nernst coefficient of the parent compound, this strong anomaly suggests the onset of orbital or SDW ordering at $T\sim160$~K. Note, however, that the magnitude of the negative contribution that appears at this temperature is about two orders of magnitude smaller than that in the parent compound, i.e., the overall effect is much more subtle.
Thus, both resistivity and Nernst effect reveal that a remnant feature of the nematic order is still governing the physics of this compound.
For completeness we mention the further upturn of $\nu(T)$  at lower temperature due to the onset of vortex motion in the superconducting state \cite{Kondrat2011}. It will not be discussed further in this paper as it solely focuses on the normal state properties.

In order to elucidate to which degree magnetism is affected, we plot \slrrt\ in Fig.~\ref{all_five}c) \cite{Hammerath2013}.  \slrrt exhibits a weak positive slope at high temperature $T\gtrsim 200$~K and approaches a linear increase (dashed line in Figs.~\ref{rho_undoped}c), \ref{all_five}c), and \ref{all_ten}c)) \cite{Hammerath2013}. This behaviour resembles the generic linear of the static susceptibility \cite{Klingeler2010} that has been heavily debated in the past \cite{Grafe2009,Grafe2008,Klingeler2010,Korshunov2008,Korshunov2009,Zhang2009b,Berciu2009,Sawatzky2009,Skornyakov2011,Chaloupka2013}, but is not at focus here.
However, at lower temperature \slrrt strongly deviates from this behavior and develops a strong increase with decreasing $T$ until it starts to decrease again after reaching a peak at about 28~K. Note that the magnitude of the increase is about one order of magnitude smaller than that related to the SDW ordering in the parent compound. The observed strong increase of \slrrt signals the progressive slowing-down of spin fluctuations of the conduction electrons \cite{Hammerath2013}.
These data show that the spin dynamics at this doping level,  in consistency with M\"o{\ss}bauer and muon spin rotation ({\textmu}SR) results \cite{Luetkens2009}, are close to but never reach long range magnetic order.

The thermal expansion at this doping level (not shown, see Fig.~2 of Ref.~\cite{Wang2009a})  as well as the  orthorhombic splitting in diffraction experiments \cite{Luetkens2009,Kondrat2009,Qureshi2010} do not show any anomaly around 160~K. The 
global
lattice degrees of freedom are therefore apparently unaffected by the remnant nematic ordering,
i.e., a long-range structural influence is absent.
Thus, the well-defined anomalies in the resistivity and the Nernst coefficient signal some form of orbital ordering,
which must be of short-range nature because otherwise a global response in the lattice degrees of freedom would be inevitable.

\subsection{Optimally doped  LaO$_{1-x}$F$_x$FeAs}

\begin{figure}[t]
\includegraphics[width=\columnwidth]{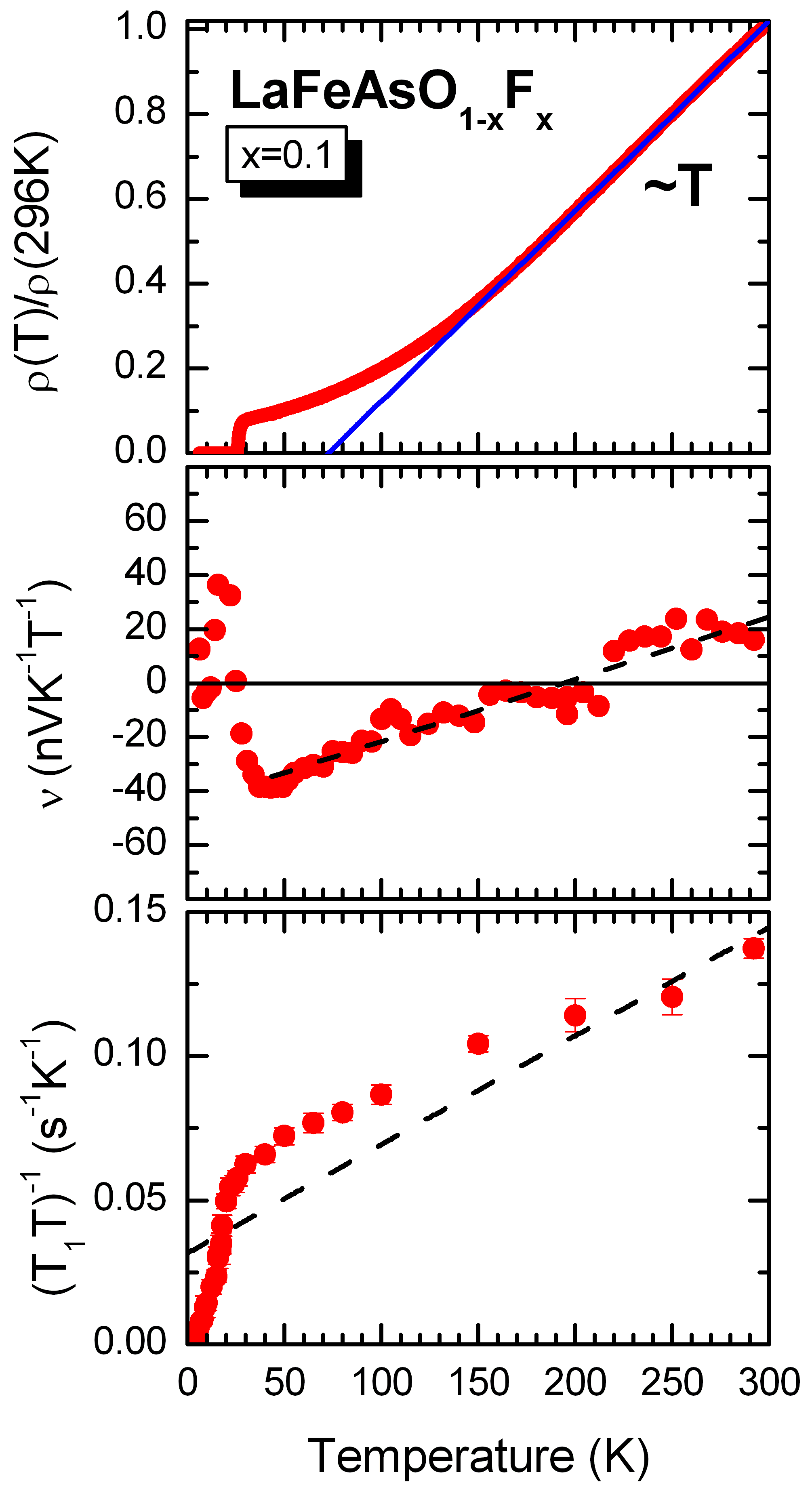}
\caption{Normalized resistivity $\rho(T)$ \cite{Hess2009} (a), Nernst coefficient $\nu(T)$ \cite{Kondrat2011} (b), and $^{75}$As-NMR spin-lattice relaxation rate \slrrt \cite{Hammerath2013,Grafe2008} of \laf at $x=0.1$. Solid and dashed lines in (a) and (b) guides to the eye. Dashed line in (c): see text.} \label{all_ten}
\end{figure}

We point out that the salient features of the remnant orbital ordering in the transport properties as well as the signature of progressive slowing-down of spin fluctuations as evidenced by the nuclear spin relaxation persist also up to higher doping levels ($x=0.075$) \cite{Hess2009,Hammerath2013} and vanish only close to optimal doping $x\approx0.1$.
Figure~\ref{all_ten} shows our results for the resistivity $\rho(T)$ \cite{Hess2009}, the Nernst effect $\nu(T)$ \cite{Kondrat2011} and the spin-lattice relaxation rate \slrrt \cite{Hammerath2013} of \laf at $x=0.1$, which is close to optimal doping. In the resistivity, see Fig.~\ref{all_ten}a), the temperature dependence is completely different as compared to the underdoped levels. $\rho(T)\sim T^2$ is found for $T\lesssim150$~K which smoothly develops into a linear temperature dependence at higher temperature. Remarkably, the drop-like anomaly around 160~K and the low-temperature upturn have completely disappeared \cite{Hess2009}.

Despite a similar magnitude as for $x=0.05$, the normal state Nernst coefficient as shown in Fig.~\ref{all_ten}b) \cite{Kondrat2011} displays a completely different normal-state behavior as compared to the underdoped
compound. At $T\gtrsim 40$~K, $\nu(T)$ is just roughly linear with a weak positive slope. In particular, no anomaly at about 160~K is present.

Figure~\ref{all_ten}c) shows \slrrt at $x=0.1$. The low-temperature increase of \slrrt has vanished completely at this doping level and one observes a monotonic increase which approaches the same linear increase as in Fig.~\ref{all_five} \cite{Hammerath2013,Grafe2008}. However, the deviation from such a linear increase, if any, is very small. This implies the essential absence of slow spin fluctuations at this doping level. Similarly to the underdoped case at $x=0.05$, no anomaly is present in the thermal expansion \cite{Wang2009a} and x-ray diffraction \cite{Kondrat2009}.

\section{Discussion}
\begin{figure}[ht]

\includegraphics[width=\columnwidth]{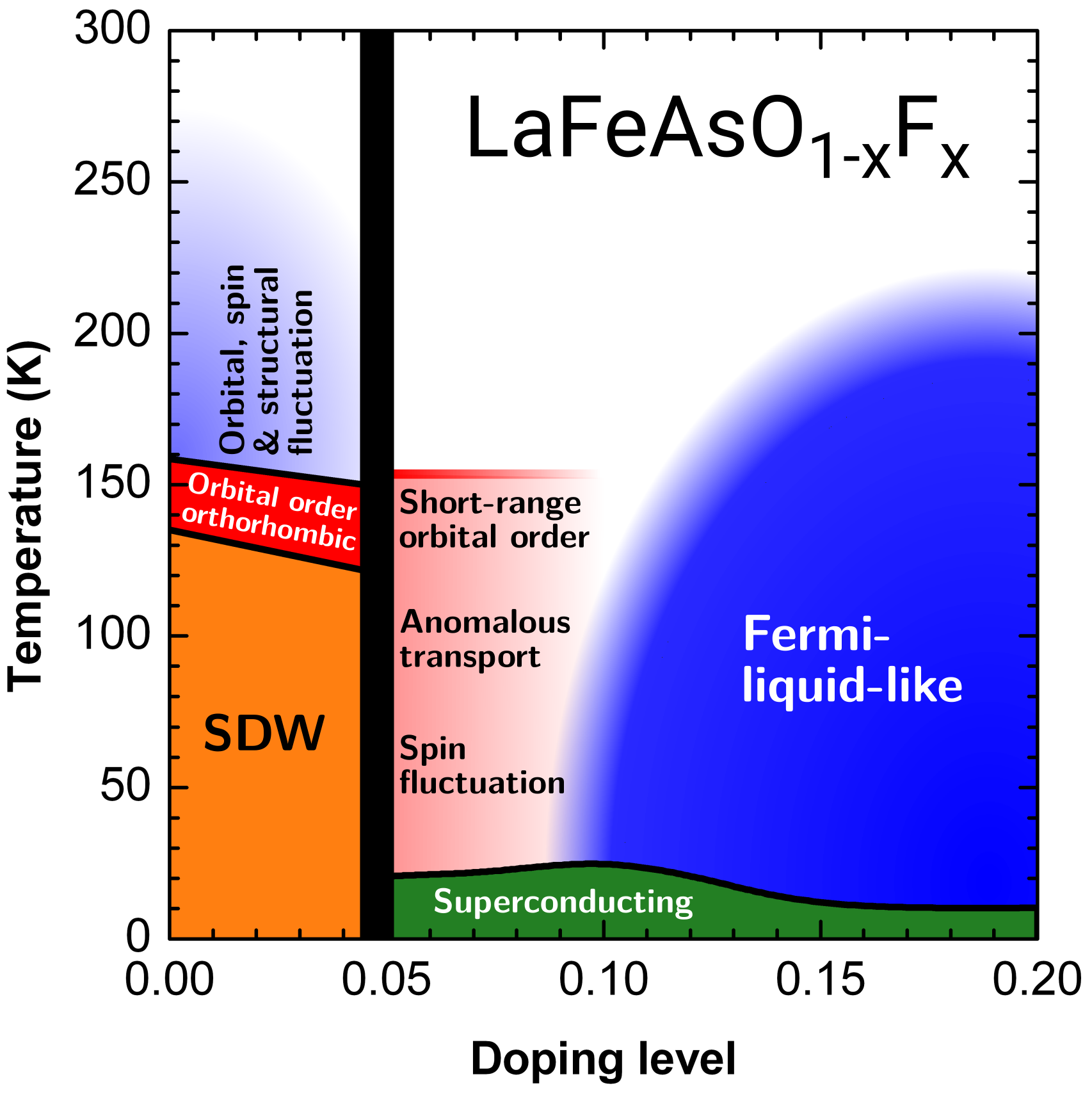}

\caption{Schematic electronic phase diagram of \laf with three distinct doping regimes: I.) Small F-concentrations $x<0.05$ including the parent compound, with long-range nematic order at $T\lesssim160$~K. II.) the underdoped superconducting regime $0.05\gtrsim x<0.1$ with short-range orbital ordering (or a slowly fluctuating variant) at $T\lesssim160$~K and progressively slowing down spin fluctuations. III.) Optimal and higher doping levels $x\gtrsim0.1$ without nematicity and concomitantly highest $T_c$, as well as Fermi-liquid-like resistivity behavior. The shaded bar at $x\approx 0.05$ indicates the crossover region between regimes I and II. } \label{phasediagram}

\end{figure}

Our main observation  derives from the data in Fig.~\ref{all_five}, viz. the apparent occurrence of  some form of orbital order in underdoped \laf at approximately the same temperature as in the parent compound while magnetic ordering is clearly absent. We recall the rationale which leads to this conclusion:
The clear anomalies in the resistivity and the Nernst coefficient in Fig.~\ref{all_five} at about 160~K signal a decrease of the transport relaxation rate (and/or the effective electron mass) and, more generally, a static Fermi surface reconstruction.
A priori, these can be connected with static orbital ordering, spin ordering, or a phonon-driven orthorhombic distortion. 
However, the latter two are not good candidates for the primary origin of the transport anomalies:
Firstly, all magnetic probes, i.e., {\textmu}SR, and M\"o{\ss}bauer spectroscopy prove the absence of static magnetic order at about 160~K \cite{Luetkens2009,Klingeler2010}, whereas the here presented NMR data of the $^{75}$As \slrrt reveal only progressively slowing down spin fluctuations with a relatively small magnitude.
Secondly, a 
long-range orthorhombic
distortion is absent as revealed by diffraction experiments and thermal expansion data \cite{Kondrat2009,Wang2009a,Qureshi2010,Luetkens2008}. 
Note, that both magnetic and structural probes do not show any anomaly around  $160$~K.
Thus, it is orbital-driven nematicity which remains as the 
most plausible
possibility for the origin of the transport anomalies.

This inferred orbital nematicy evidently is very subtle, because the transport anomalies are much less pronounced as compared to the parent compound \lao where all degrees of freedom, viz. orbitals, spins and the lattice orthorhombic distortion participate in the ordering. This subtlety implies short-range orbital ordering or a slowly fluctuating variant, with length and time scales that are comparable to the electronic mean free path respectively the transport life time.

If one considers the doping evolution of orbital ordering with the data at hand the following scenario is apparent: In the parent compound, i.e., in \laf at $x=0$, long-range orbital order develops at $T_s=160$~K, with an extended regime of fast fluctuations at higher  temperatures. As this orbital ordering is long-range, the lattice can follow the orbital structure and the long-range orthorhombic distortion concomitantly develops. Long-range antiferromagnetic order then establishes at somewhat lower temperature at $T_N$, permitted by the broken $C_4$ symmetry. At underdoping ($x=0.05$), short-range orbital order sets in at about 160~K, i.e., at about the same temperature where long-range orbital order develops in the undoped case. It is thinkable,  that the orbital ordering is only quasi-static, i.e., the hort-range orbital ordering is slowly fluctuating with a time scale that is much slower than the fast fluctuations in the parent compound at $T>160$~K.
In any case, the $C_4$ symmetry is only broken on a local scale. In this situation the lattice 
can at most follow on the local scale, i.e., a long-range orthorhombic distortion is impossible. However,
the emergence of orthorhombically  distorted nanoscale domains which are induced by the orbital 
structure but are invisible to the global lattice probes cannot be excluded. 
In an analogous way the long-range magnetic order is prevented. Nevertheless, the short-range orbital order
fosters the slowing-down of spin fluctuations. Note, that a fluctuation regime at $T>160$~K of the short-range orbital ordering is not recognizable from the transport data, which implies that the overall amplitude of the ordering and of its fluctuations is relatively small. Finally, orbital ordering, even of short range nature, is apparently completely absent at the near optimal doping level $x=0.1$ as no anomalies occur at about 160~K.

The actual origin for the sketched doping evolution of nematicity in \laf remains unclear. One might speculate that the orbital ordering becomes increasingly unfavorable upon electronic band filling with doping. This scenario seems, however, rather unlikely because on a local scale the energetics of the orbital ordering obviously stay the same upon doping. This derives from the fact that the actual temperature where the ordering sets in is the same for $x=0$ and $x=0.05$.
Alternatively, one might conclude from the evolution from long-range to short-range ordering a spatially inhomogenous electronic structure due to the doped charges. In such a scenario, orbital ordering could occur locally in sample areas which electronically resemble much the parent compound and would be quenched in the direct vicinity of a doped charge. The transition from long range ordering to short range ordering as a function of doping can then be understood in terms of a percolation threshold which separates the electronic phase diagram into a low-doping regime with long-range nematic order without superconductivity and an underdoped regime with quenched, i.e., short-range (or fluctuating) nematic order which allows superconductivity. Remarkably, optimal superconductivity is reached when no signature of the short-range nematic order is present. If one stays in the sketched scenario it is plausible that a total quenching of nematicity is reached at about 10\% doped charges where distance between doped charges is already very small.
We point out that this picture of the evolution of nematicity in \laf is supported by clear-cut experimental evidence from NQR-measurements \cite{Lang2010,Lang2015} for two microscopically distinct charge environments at the As nucleus for $x<0.1$, with one corresponding to that of the parent compound and the other to that of higher-doped compounds. It has also been suggested based on NQR data that percolation would indeed control the occurence of orthorhombicity and static magnetism \cite{Lang2015}.

We summarize our qualitative findings in a schematic phase diagram of \laf which is shown in Fig.~\ref{phasediagram} where we include earlier comprehensive results from resistivity \cite{Hess2009}. The parent compound and non-superconducting compounds at low doping levels exhibit a long-range ordered orbital-driven nematic state at low temperature. At higher temperature, i.e., prior to the transition to the nematic state, orbital, structural and magnetic fluctuations govern the physics of the material. This situation persists at small doping levels where the nematic order is somewhat destabilized due to the doped charges but remains  long-range. Above a critical doping value $x<0.05$, the long-range nematic order state is completely suppressed, presumably due to an exceeded percolation limit which quenches the long-range order, and superconductivity occurs. The clear anomalies in the resistivity and the Nernst coefficient at about 160~K and the absence of a structural distortion and of magnetic order provide the evidence for short-range orbital order  (which might also be slowly fluctuating) at the same temperature as the long-range order in the parent compound. 
Naturally, due to the inevitable coupling of orbital and lattice degrees of freedom, one can expect nanoscale orthorhombically distorted lattice domains, which are induced by the orbital ordering.
At the same time, this phase possesses signatures of incipient magnetic order, evidenced by slow spin fluctuations.
This short-range or fluctuating nematic state fades out at about optimal doping where the resistivity approaches a less unconventional normal state with a $\sim T^2$ dependence at low temperature, signaling Fermi-liquid-like behavior and a linear temperature dependence at higher temperatures.

It is interesting to note that the orbital driven nematic order in \laf bears some resemblance with the recently reported nematicity in FeSe \cite{Baek2014,Boehmer2015}, in the sense that the long-range orbital ordering, accompanied by an orthorhombic distortion, occurs without magnetic order, with the difference that in \lao long-range magnetic order sets in just a few Kelvin below whereas it remains absent in FeSe. 
It is also worth to mention that the doping evolution of nematic fluctuations in \laf is quite different from that in Co- and K-doped \bfa. In \laf short-range orbital order (or fluctuations of it) seem to disappear towards optimal doping whereas in \bkfa and \bfca nematic fluctuations have been reported to be present over the entire superconducting dome \cite{Boehmer2014a} with a tight connection between magnetic and lattice fluctuations \cite{Fernandes2013}. From these findings for the \bfa-based compounds it has been conjectured the nematic fluctuations could be involved in the superconducting pairing. Our finding of optimal superconductivity with concomitantly disappearing nematic fluctuations renders this possibility less likely for \laf, and suggest that the nematic phase actually competes with superconductivity.


\section{Summary}
We have investigated the iron-oxypnictide superconductor \laf in the normal state with a focus on signatures of nematic phases in resistivity, Nernst effect, and $^{75}$As NMR data. For undoped \lao, very pronounced anomalies in all these quantities signal the onset of long-range nematic order. At the underdoped level $x=0.05$ much weaker but still distinct anomalies at about the same temperature where undoped \lao develops long-range nematic ordering remain present in the transport coefficients. Despite these signatures of a nematic state, a detectable orthorhombic lattice distortion and magnetic order are absent, and the $^{75}$As-NMR spin-lattice relaxation rate \slrrt reveals a progressive slowing down of spin fluctuations. We conclude from these observations that short-range or slowly fluctuating orbital ordering sets in near 160~K, which is the onset temperature of long-range nematic order in the parent compound \lao. All anomalies in the transport coefficients and also the indications of slow spin fluctuations are absent at $x=0.1$, i.e., near optimal doping. The results therefore suggest that the nematic phase competes with superconductivity as it is gradually destroyed by doping and vanishes upon reaching a maximum critical temperature at optimal doping.


\section{Acknowledgement}

This work has been supported by the Deutsche Forschungsgemeinschaft (DFG) through the Priority Programme SPP1458 (Grant BE1749/12 and GR3330/2).


%
%
%

\begin{thebibliography}{[10]}
\bibitem{dila}{The samples are fixed by a small force in the capacitance dilatometer resulting in a pressure of the order of $\approx 100$~kPa applied along the direction of the measured length changes.}
\end{thebibliography}

\providecommand{\WileyBibTextsc}{}
\let\textsc\WileyBibTextsc
\providecommand{\othercit}{}
\providecommand{\jr}[1]{#1}
\providecommand{\etal}{~et~al.}

\end{document}